\documentclass{PoS}

\newcommand{\comment}[1]{}

\newcommand{\lr}[1]{ \left( #1 \right) }
\newcommand{\lrs}[1]{ \left[ #1 \right] }

\newcommand{\vev}[1]{ \langle \, #1 \, \rangle }

\title{Competing order in the fermionic Hubbard model on the hexagonal graphene lattice}

\ShortTitle{Fermionic Hubbard model on the hexagonal graphene lattice}

\author{Pavel Buividovich\\
        Institut f\"ur Theoretische Physik, Universit\"at Regensburg, 93053 Regensburg, Germany\\
        E-mail: \email{pavel.buividovich@physik.uni-regensburg.de}}

\author{Dominik Smith\\
        Institut f\"ur Theoretische Physik, Justus-Liebig-Universit\"at Gie\ss en, 35392 Gie\ss en, Germany\\
       E-mail: \email{dominik.smith@theo.physik.uni-giessen.de}}

\author{Maksim Ulybyshev\\
        Institut f\"ur Theoretische Physik, Universit\"at Regensburg, 93053 Regensburg, Germany\\
        E-mail: \email{maksim.ulybyshev@physik.uni-regensburg.de}}

\author{\speaker{Lorenz von Smekal}\thanks{This work was supported by DFG grants BU 2626/2-1 and SM 70/3-1.}\\
        Institut f\"ur Theoretische Physik, Justus-Liebig-Universit\"at Gie\ss en, 35392 Gie\ss en, Germany\\
       E-mail: \email{lorenz.smekal@theo.physik.uni-giessen.de}}

\abstract{We study the phase diagram of the fermionic Hubbard model on
  the hexagonal lattice in the space of on-site and nearest neighbor
  couplings with Hybrid-Monte-Carlo simulations. With pure on-site
  repulsion this allows to determine the critical coupling strength
  for spin-density wave formation with the standard approach of
  introducing a small mass term, explicitly breaking the sublattice
  symmetry. The analogous mass term for charge-density wave formation
  above a critical nearest-neighbor repulsion, on the other hand,
  would introduce a fermion sign problem. The competition between the
  two and the phase diagram in the space of the two coouplings
  can however be studied in simulations without explicit sublattice
  symmetry breaking. Our results compare qualitatively well with the
  Hartree-Fock phase diagram. We furthermore demonstrate how
  spin-symmetry breaking by the Euclidean time discretization
  can be avoided also, when using an improved fermion action based on
  an exponetial transfer matrix with exact sublattice symmetry.}

\FullConference{34th annual International Symposium on Lattice Field Theory\\
         24-30 July 2016\\
         University of Southampton, UK}

\graphicspath{{./figures/}}

\begin{document}
\sloppy

\section{Introduction}
\label{sec:intro}

Possible insulator-semimetal Mott transitions in the hexagonal Hubbard model with varying on-site repulsion and nearest- and next-to-nearest-neighbor interaction strengths have been a subject of active theoretical studies which led to a rather rich phase diagram with anti-ferromagnetic (AF) spin-density wave (SDW) and charge-density wave (CDW) phases \cite{Sorella:1992,Semenoff:84:1,Herbut:06:1}, topological insulators \cite{Raghu:07:1} and spontaneous Kekul\'e distortions \cite{PhysRevLett.98.186809}. While by now it is well established both experimentally \cite{Elias:12:1} and numerically \cite{Buividovich:13:5,Smekal:13:3} that free suspended graphene is a semimetal, the interest in this rich phase structure is still not purely academic. It is expected that some of these phases can be realized in mechanically strained graphene \cite{Assaad:15:1}, in other recently discovered 2D materials with hexagonal lattices such as phosphorene \cite{Liu:14:1}, silicene and germanene \cite{Cahangirov:09:1} or by designing an ``artificial graphene'' with optical lattices \cite{nature10871}.

While semi-analytic methods such as a large-$N$ renormalization group
fixed-point analysis \cite{Herbut:06:1} or a variational Hamiltonian
approach \cite{Semenoff:12:1}, which is equivalent to solving
the Dyson-Schwinger equations on the hexagonal lattice in
Hartree-Fock approximation \cite{Kleeberg:2016},
%\cite{Semenoff:12:1,Smekal:13:2,Smekal:16:1}
provide a reasonably good qualitative description of the phase diagram
in the parameter space of on-site and nearest-neighbor couplings,
there are considerable variations in the actual values of critical couplings
and the locations of phase-transitions among these methods. In
order to produce quantitative results for the phase diagram one should
therefore turn to first-principles numerical simulations. The
transition separating semimetal and SDW
phases in the hexagonal Hubbard model with only on-site interaction
has recently been studied in detail in \cite{Assaad:13:1} using
Determinantal Quantum Monte-Carlo with ground-state
projection. Here we present first results for the phase diagram of
this model with both on-site and nearest-neighbor interactions from
Hybrid Monte-Carlo (HMC) simulations. In contrast to Determinantal
Quantum Monte-Carlo, this algorithm avoids the use of explicitly
calculated ratios of fermionic determinants in the Metropolis
accept-reject step which significantly speeds up the sampling for
large lattice sizes \cite{Buividovich:16:1}. We also demonstrate how to
perform simulations with exact sublattice symmetry, by
using a gapless single-particle Hamiltonian and a novel discretization
of the fermionic action in Euclidean time. In particular, this allows us
to study CDW order and its competition with SDW order in the space of
on-site and nearest-neighbor couplings.

\section{HMC simulations in the chiral limit of graphene}
\label{sec:chiral_limit}

Since spontaneous symmetry breaking cannot happen in a system with a
finite number of degrees of freedom, detecting it in Monte-Carlo
simulations on a finite lattice is a nontrivial numerical problem. The
standard strategy also used in our previous HMC studies of the
semimetal-insulator transition in suspended graphene
\cite{Buividovich:13:5,Smekal:13:3} is to introduce a small explicit
mass term $m_{SDW}$ which induces SDW order and thus opens a small gap
in the energy spectrum of the single-particle Hamiltonian. The
measured SDW order parameter is then extrapolated to infinite volume
and $m_{SDW} \to 0$ in the right order, yielding a
nonzero result in the phase where the sublattice symmetry is spontaneously
broken by the anti-ferromagnetic SDW ground state.

\begin{figure}[t]
%\begin{center}
\includegraphics[width=1.\linewidth]{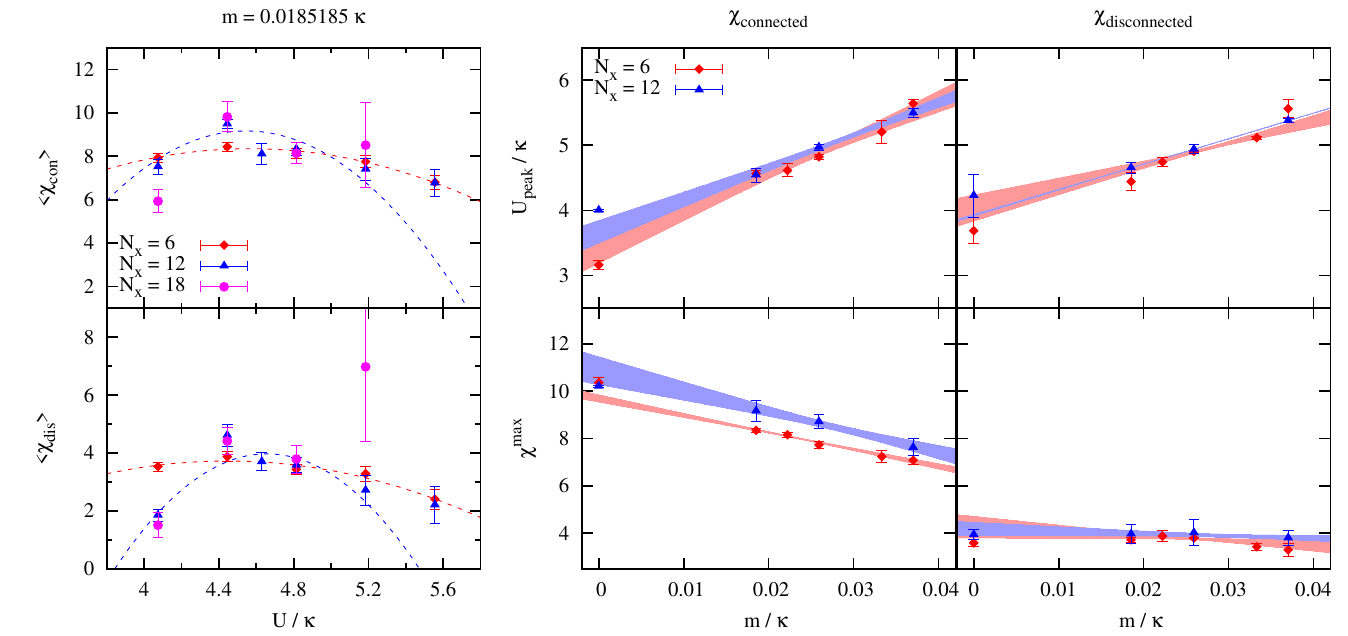}
%\end{center}
\caption{Dependence of connected and disconnected parts of the
  susceptibility for SDW order-parameter fluctuations on the
  on-site interaction strength at small but finite mass
  with interpolations to extract the
  peaks (left) which are then extrapolated and compared with results
  from simulations at exactly zero mass (right).}
\label{fig:susceptibility_extrapolation}
\end{figure}

One of the disadvantages of this method is the need to perform two
extrapolations (towards infinite volume and zero mass). Moreover, the
corresponding mass term to study CDW order violates the particle-hole
symmetry and thus leads to a fermion-sign problem in HMC simulations. An
alternative strategy, for which only the infinite-volume
extrapolation is required, is to study the susceptibilities of
the relevant order parameters at exactly zero mass. Their peaks can
indicate phase transitions even when the latter vanish in
a finite volume. To test this, we have verified numerically that the
zero-mass extrapolations of the susceptibilities agree with the
results of simulating these susceptibilities at exactly zero mass on a
finite lattice within errors. The comparison is limited to the case of
only on-site interactions of strength $U$ where spin-density wave
formation is well established for $U> U_c$
\cite{Sorella:1992,Assaad:13:1}, and simulations with the relevant
mass term do not suffer from a sign problem. In the left panel of
Fig.~\ref{fig:susceptibility_extrapolation} we show the
dependence of the susceptibility $\chi$ of the SDW order parameter on
$U$ for the smallest mass we have used in our simulations. We
separately plot the contributions of connected ($\chi_\mathrm{con}$) and
disconnected ($\chi_\mathrm{dis}$) fermionic diagrams to the susceptibility
at different lattice volumes. Both contributions feature clearly
visible peaks which indicate the transition point. There is some
evidence for a volume scaling indicative of a phase transition, but
better statistics on larger lattices will be necessary to confirm
that. In the right panel we compare the results of linear
extrapolations of positions and heights of the susceptibility peaks
with the results obtained directly at zero mass. In all cases the
latter are close to the $95 \%$ confidence intervals for linear
extrapolations. There is a considerable shift of the susceptibility
peaks towards smaller values of $U$ as the explicit SDW mass is
reduced, but both, the extrapolated and the measured peak positions at
zero mass are overall consistent with previous determinations of the
critical coupling $U_c \simeq 3.8 \kappa $ \cite{Assaad:13:1} in units
of the tight-binding hopping parameter $ \kappa$.

Finally we notice that in contrast to expectations from lattice QCD,
in practice our HMC simulations run quite smoothly even at exactly
zero mass. They do not get stuck in some region of
configuration space (such as fixed-topology sectors in QCD). The
reason most likely is that there is simply no non-trivial topology for
the Hubbard fields involved in graphene simulations, and the geometry
of manifolds on which the fermion determinant is zero is certainly
quite different from that in QCD. For simple few-site tight-binding
models in the large-temperature limit we have checked explicitly
that zeros of the fermion determinant form manifolds of dimension
$N-2$ in the $N$-dimensional configuration space of the Hubbard field. It
is clear that manifolds of such dimension cannot enclose an
$N$-dimensional region where the molecular dynamics
could get stuck. In rare events it can get close to one of these
$N-2$-dimensional manifolds which only results in a large and positive
molecular-dynamics energy difference, however, and therefore leads to
rejection in the Metropolis check.

The possibility of simulations at exactly zero mass can be proven in a
more rigorous way by noticing that for lattice sizes which are not
multiples of three Dirac points are not within the discrete set of
lattice momenta. As a result, the spectrum of the single-particle
Hamiltonian on such lattices has a small gap inversely proportional to
the lattice size, and the fermion determinant never crosses zero. Thus
there are no impenetrable barriers in the HMC dynamics, and the whole
configuration space can be efficiently sampled. We call this
geometry-dependent energy gap a ``geometric mass'' term. In contrast
to the mass terms which explicitly break sublattice symmetry and
induce spin-density or charge-density wave order, the geometric mass
term does not break sublattice symmetry and does not favor any order
in the ground state. It is thus ideally suited for or purposes.

\section{Phase diagram with on-site and nearest-neighbor interactions}
\label{sec:phase_diagram}

To identify the phase transitions in the space of combined on-site $U$ and
nearest-neighbor repulsion of strength $V$ we follow the approach
of \cite{Herbut:14:1} and study the volume dependence of expectation
values of sums of squares of sublattice averages as order parameters.
SDW order, for example, can be identified from averages of
squared sublattice magnetizations
\begin{eqnarray}
\label{order_parameter_sq}
  \vev{S^{i}}  = \frac{1} {L^2} \sqrt{ \langle {\left( \sum_{x=(A,\xi)} \hat{S}_{x}^{i} \right) }^2 \rangle + \langle {\left( \sum_{x=(B,\xi)} \hat{S}_{x}^{i} \right) }^2 \rangle } ,
\end{eqnarray}
where $L$ is the lattice size, $\hat S^i_{x} = \frac{1}{2} ( \hat
a^\dag_{x, \uparrow} , \, \hat a^\dag_{x, \downarrow} ) \sigma_i
\left({\hat a_{x, \uparrow}} \atop {\hat a_{x, \downarrow}}  \right)$
is the spin operator at lattice site $x = (\alpha, \xi)$, where
$\alpha=A,B$  is sublattice index and $\xi$ the elementary cell
coordinate, and $\sigma_i$ are the Pauli matrices. The sum over both
sublattices is used to reduce statistical errors. This observable can
be easily expressed in terms of fermionic two-point Greens
functions. An analogous definition of $\vev{Q}$ as (the square root
of) the sum of the expectation values of squared electric charges per
sublattice can be used to study CDW order at half-filling.

\begin{figure}[t]
\begin{center}
\includegraphics[width=1.\linewidth]{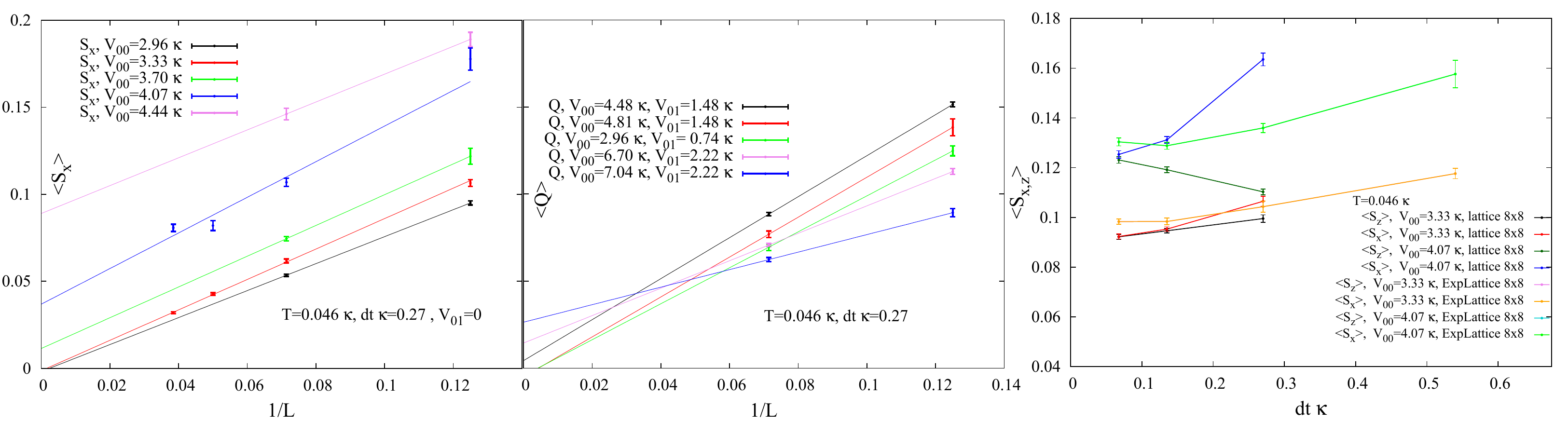}
\end{center}
\caption{Infinite volume extrapolations of $\vev{S^x}$ at vanishing
  nearest-neighbor coupling $V_{01}\equiv V$ indicating SDW order for $
   V_{00}\equiv U   \geq U_c \sim  3.7 \kappa $ (left), and of
   $\vev{Q}$ for various combinations of $U$ and $V$ (middle);
   comparison of $\vev{S^x}$ and $\vev{S^z}$
   for different Euclidean time discretizations (right), leading to
   rather strong spin-symmetry breaking in the conventional
   local fermion action whereas the results for $\vev{S^x}$ and
   $\vev{S^z}$ completely coincide with the improved non-local action
   (\protect\ref{exp_improved_action}) with exact sublattice symmetry. }
\label{fig:example-scaling}
\end{figure}

Observables as in Eq.~(\ref{order_parameter_sq}) can have finite $L
\rightarrow \infty $ limits only in phases with the corresponding
long-range order. The volume dependence of $\vev{S^x}$ in our geometric mass
simulations, i.e.~without explicit sublattice symmetry breaking, is
shown in the leftmost panel of Fig.~\ref{fig:example-scaling} for
selcted values of the on-site coupling $U \equiv V_{00}$ at vanishing
$V\equiv V_{01}$.  The transition to the AF phase with SDW order is
again observed at around $U_c \simeq 3.7$, in agreement with the
previous section. The middle panel of  Fig.~\ref{fig:example-scaling}
shows selected examples with and without CDW order as signaled by a
non-zero extrapolation of $\vev{Q}$ depending on the strength of the
nearest-neighbor repulsion. We have performed analogous simulations
with zero mass for a set of points covering the most interesting
region of the phase diagram and used infinite-volume extrapolations of
$\vev{S^i}$ and $\vev{Q}$ to identify spontaneous symmetry breaking
with SDW or CDW order.

\begin{figure}[t]
  \includegraphics[width=0.524\linewidth]{{{phase_diag}}}
\hfill\includegraphics[width=0.47\linewidth]{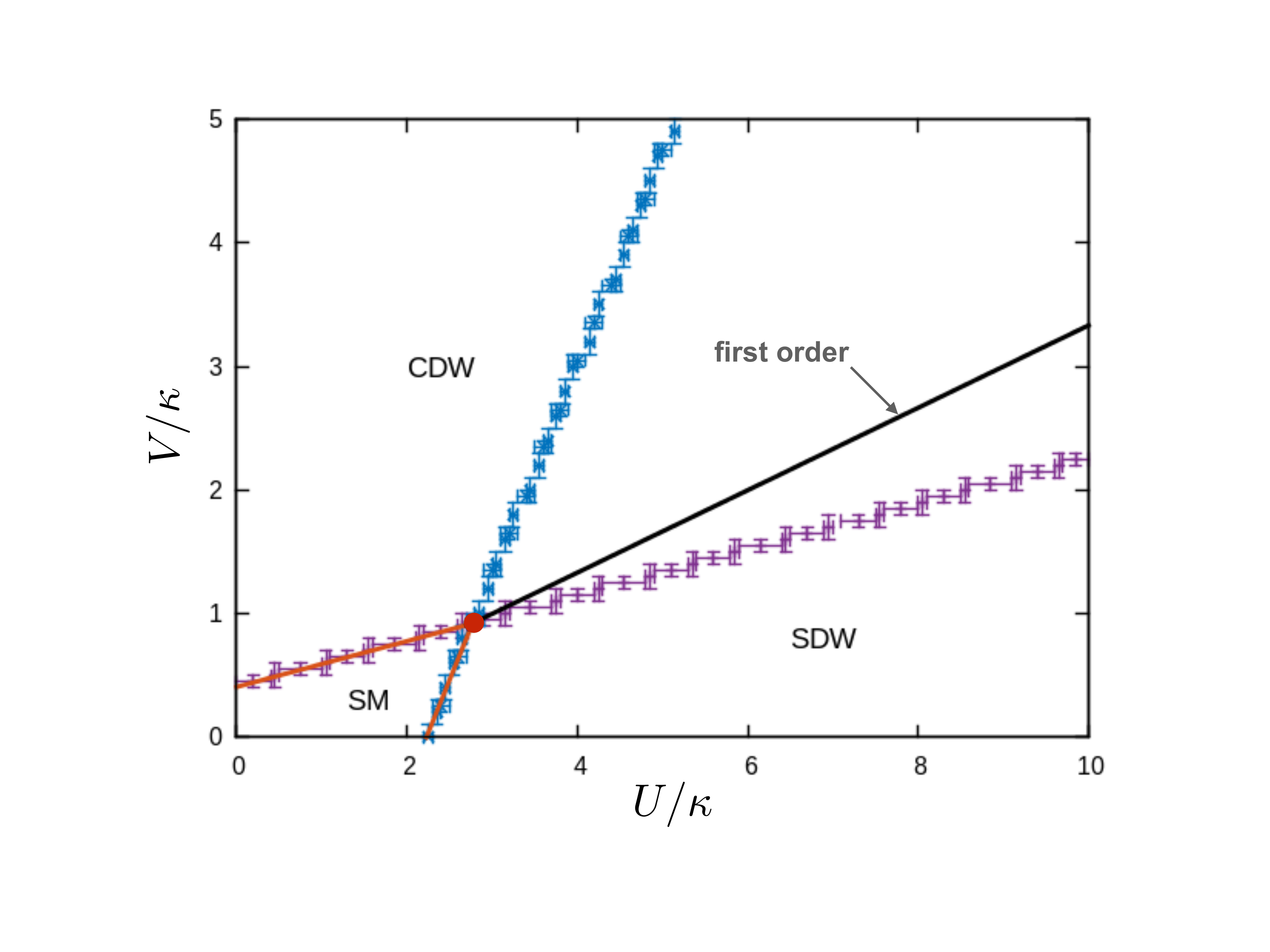}
    \caption{Preliminary phase diagram of the Hubbard model on the
      hexagonal lattice in the parameter space of on-site $U$ and
      nearest-neighbor interaction $V$ from the HMC simulations with
      geometric mass of this work (left) compared to the Hartree-Fock
      phase diagram  (right) from Dyson-Schwinger equations
      \cite{Kleeberg:2016}.}
    \label{fig:phase_diag}
\end{figure}

The resulting numerical estimate of the phase diagram is shown in the
left panel of Fig.~\ref{fig:phase_diag} where red squares and blue
triangles mark non-zero infinite-volume extrapolations
of the SDW and CDW order parameters, respectively. The black dots mark
the points where both extrapolations are consistent with zero
as evidence for the semimetal phase. The straight yellow
line separates the region with $V > U/3$ where our HMC algorithm can no-longer
be used because the interaction matrix ceases to be positive definite
thus invalidating the Hubbard-Stratonovich transfromation
\cite{Buividovich:16:1}. Along this line with $V=U/3$, due to the
coordination number three of the graphene lattice, SDW and CDW
ground states have equal interaction energies and one therefore
expects this to become a first order transition line in the
strong-coupling limit. This is also seen in our Hartree-Fock (HF) phase
diagram from the Dyson-Schwinger equations on the hexagonal lattice
\cite{Kleeberg:2016} which agrees with that of Ref.~\cite{Semenoff:12:1}
and is shown in the right panel of Fig.~\ref{fig:phase_diag}. The black
first-order transition line in the strong coupling-region of the HF
phase diagram ends where the SDW, CDW semimetal phases meet, in what is
probably a Lifshitz point (marked in red). The boundary lines of the
semimetal (SM) region both mark continuous transitions, most likely in the
Gross-Neveu universality class as predicted for the SDW transition
\cite{Herbut:06:1,Assaad:13:1}. They continue at larger couplings as
spinodal lines limiting possible coexistence regions around the
first-order line. Interestingly, for the two points of our data close
to this line with the strongest couplings both, the SDW and
the CDW order parameter extrapolate to non-zero infinite-volume limits
indicating that they might indeed be in the co-existence
region. At the moment our HMC data does not allow to determine the
order of the transitions between semimetal, SDW and CDW phases.
Overall, the phase diagram agrees qualitatively quite well with the
expectations from the HF calculation, however.

%\section{Violation of symmetry between different spin components due
%to time discretization and the way to cure it}
\section{Spin-symmetry violation by time discretization and improved
  fermion action}
\label{sec:sublat_symm_break}

In absence of spin-dependent interactions, the two physical spin
states of the electrons on the hexagonal lattice are treated as two
independent fermion fields which can be freely rotated into each other
by unitary $U(2)$ transformations analogous to different flavors in
QCD. This implies that the squares of all spin components in
(\ref{order_parameter_sq}) should be equal to one another. In
practice, however, one observes that $\vev{S^x}$
is significantly larger than $\vev{S^z}$, as seen in the right panel
of Fig.~\ref{fig:example-scaling} (without the ``Exp'' in the label),
and that the difference vanishes only very slowly in the contimuum
limit. We have simulated lattices with up to $N_\tau = 640$ time
slices to verify that the spin-symmetry does  get restored when
approaching the continuum limit in the Euclidean time discretization,
eventually.

The origin of this asymmetry lies in an identification of physical
say spin-up electron states with particles and spin-down with
anti-particles which, on one hand, depends on the choice of the basis
and, on the other hand, involves a sublattice-dependent phase
\cite{Buividovich:13:5,Smekal:13:3}. This couples spin and sublattice
symmetries. In terms of the single-particle Hilbert space one can
define a generator $\Sigma_{xy}$ of sublattice symmetry whose
non-vanishing matrix elements for  $x = y$
are $+1$ on one sublattice and $-1$ on the other sublattice. This
generator is completely analogous to $\gamma_5$ for the Dirac
Hamiltonian. In particular, the single-particle Hamiltonian satisfies
(in the absence of SDW or CDW mass terms) the identity $\Sigma h
\Sigma = - h$. Thus exchanging the two sublattices is
equivalent to exchanging positive- and negative-energy states, which
are equivalent by virtue of the particle-hole symmetry of the bipartite
lattice. At the level of the finite-temperature partition function,
this identity implies $\Sigma e^{-\beta h} \Sigma = e^{\beta
  h}$. However, if one discretizes the Euclidean time into intervals
of size ${\Delta \tau} = \beta/N_{\tau}$ and replaces the exponent
$e^{-\beta h}$ by $\lr{1 - {\Delta \tau} h}^{N_{\tau}}$, such an
identity no-longer holds, since $\Sigma \lr{1 - {\Delta \tau}
  h}^{N_{\tau}} \Sigma = \lr{1 + {\Delta \tau} h}^{N_{\tau}} \neq
\lr{1 - {\Delta \tau} h}^{-N_{\tau}}$. In other words, the particle
transfer matrix is no-longer the inverse of the hole transfer matrix
after exchanging sublattices, the particle propagating backwards in
time is no-longer equivalent to a hole, and thus the combined
particle-hole and sublattice symmetries are violated by corrections of
order $\Delta \tau$. Since particles and holes were identified with
certain spin components, this violation translates into one of the
spin symmetry.

To overcome this difficulty we propose to use a spatially non-local
fermion action in which the factors $\lr{1 - {\Delta \tau} h}$ are
replaced by the exponents $e^{-{\Delta \tau} h}$,
\begin{eqnarray}
\label{exp_improved_action}
 S\lrs{\bar{\psi}_{\tau}, \psi_{\tau}, \phi_{\tau}} = \sum\limits_{\tau} \lr{\bar{\psi}_{\tau} \psi_{\tau} - \bar{\psi}_{\tau} e^{-{\Delta \tau} h} e^{i \phi_{\tau}} \psi_{\tau+\Delta \tau}} ,
\end{eqnarray}
where $\psi_{\tau}$ are the fermion fields at time slice $\tau$ (with
spatial indices omitted here), $\phi_{\tau}$ are the Hubbard fields at
time slice $\tau$ (interpreted as a diagonal matrix in the
single-particle Hilbert space) and $h$ is the free single-particle
Hamiltonian of the tight-binding model. One can show that even in the
presence of fluctuating Hubbard fields this action has an exact
sublattice-particle-hole symmetry, and thus restores the symmetry
between different spin components.

To illustrate this, in the right panel of
Fig.~\ref{fig:example-scaling} we also show the
observables  $\vev{S^i}$ from Eq.~(\ref{order_parameter_sq}) for SDW
order calculated with the improved fermion action
(\ref{exp_improved_action}). The corresponding data points (with the
``Exp'' in the label) for $\vev{S^x}$ and $\vev{S^z}$ lie
exactly on top of each other which demonstrates the complete
restoration of spin symmetry. An additional advantage is that
discretization artifacts seem to be much smaller with the improved
action (\ref{exp_improved_action}). Being non-local and invariant
under chiral sublattice symmetry, the fermion action
(\ref{exp_improved_action}) is in fact similar to the overlap action
in lattice QCD. The non-locality of the action renders the Conjugate
Gradient inversions within the HMC algorithm rather slow. We have
therefore implemented alternative inversion algorithms which
significantly accelerate HMC simulations with the
non-local action (\ref{exp_improved_action}) as will be
presented elsewhere.

In summary, we have developed the machinery to study the phase diagram
of the hexagonal Hubbard model in the space of on-site and
nearest-neighbor coupling. We have identified suitable observables to
study the competition between spin-density and charge-density wave
order without explicit sublattice-symmetry breaking, and we have
implemented an improved fermion action that does not break this
symmetry by the Euclidean time discretization either. To reduce the
systematic uncertainties we currently repeat the
calculations of the phase diagram with the improved action.

%\bibliographystyle{mybibstyle}
%\bibliography{Buividovich,Lorenz}

\begin{thebibliography}{10}
\expandafter\ifx\csname url\endcsname\relax
  \def\url#1{{\tt #1}}\fi
\expandafter\ifx\csname urlprefix\endcsname\relax\def\urlprefix{URL }\fi
\providecommand{\eprint}[2][]{\url{#2}}

\bibitem{Sorella:1992}
S.~Sorella, E.~Tosatti, \href{http://stacks.iop.org/0295-5075/19/i=8/a=007}{Europhys. Lett. {\bf 19} (1992), 699}.

\bibitem{Semenoff:84:1}
G.~W. Semenoff, \href{http://link.aps.org/doi/10.1103/PhysRevLett.53.2449}{Phys. Rev. Lett. {\bf 53} (1984), 2449 -- 2452}.

\bibitem{Herbut:06:1}
I.~F. Herbut, Phys. Rev. Lett. {\bf 97} (2006), 146401,
  \href{http://arxiv.org/abs/cond-mat/0606195}{ArXiv:cond-mat/0606195}.

\bibitem{Raghu:07:1}
S.~Raghu, X.~Qi, C.~Honerkamp, S.~Zhang, Phys. Rev. Lett. {\bf 100} (2008),
  156401, \href{http://arxiv.org/abs/0710.0030}{ArXiv:0710.0030}.

\bibitem{PhysRevLett.98.186809}
C.-Y. Hou, C.~Chamon, C.~Mudry, \href{http://link.aps.org/doi/10.1103/PhysRevLett.98.186809}{Phys. Rev. Lett. {\bf 98} (2007), 186809}.

\bibitem{Elias:12:1}
A.~S. Mayorov, D.~C. Elias, I.~S. Mukhin, S.~V. Morozov, L.~A. Ponomarenko,
  K.~S. Novoselov, A.~K. Geim, R.~V. Gorbachev, Nano Lett. {\bf 12} (2012),
  4629--4634, \href{http://arxiv.org/abs/1206.3848}{ArXiv:1206.3848}.

\bibitem{Buividovich:13:5}
M.~V. Ulybyshev, P.~V. Buividovich, M.~I. Katsnelson, M.~I. Polikarpov, Phys.
  Rev. Lett. {\bf 111} (2013), 056801, \href{http://arxiv.org/abs/1304.3660}{ArXiv:1304.3660}.

\bibitem{Smekal:13:3}
D.~Smith, L.~{von Smekal}, Phys. Rev. B {\bf 89} (2014), 195429,
  \href{http://arxiv.org/abs/1403.3620}{ArXiv:1403.3620}.

\bibitem{Assaad:15:1}
H.~Tang, E.~Laksono, J.~N.~B. Rodrigues, P.~Sengupta, F.~F. Assaad, S.~Adam,
  Phys. Rev. Lett. {\bf 115} (2015), 186602, \href{http://arxiv.org/abs/1505.04188}{ArXiv:1505.04188}.

\bibitem{Liu:14:1}
H.~Liu, A.~T. Neal, Z.~Zhu, Z.~Luo, X.~Xu, D.~Tom\'{a}nek, P.~D. Ye, \href{http://dx.doi.org/10.1021/nn501226z}{ACS Nano {\bf 8} (2014), 4033 -- 4041}.

\bibitem{Cahangirov:09:1}
S.~Cahangirov, M.~Topsakal, E.~Akt\"{u}rk, H.~\c{S}ahin, S.~Ciraci, \href{http://dx.doi.org/10.1103/PhysRevLett.102.236804}{Phys. Rev. Lett. {\bf 102} (2009), 236804}.%, \href{http://arxiv.org/abs/0811.4412}{ArXiv:0811.4412}.

\bibitem{nature10871}
L.~Tarruell, D.~Greif, T.~Uehlinger, G.~Jotzu, T.~Esslinger, \href{http://dx.doi.org/10.1038/nature10871}{Nature {\bf 483}
  (2012), 302--305}.

\bibitem{Semenoff:12:1}
Y.~Araki, G.~W. Semenoff, Phys. Rev. B {\bf 86} (2012), 121402,
  \href{http://arxiv.org/abs/1204.4531}{ArXiv:1204.4531}.

\bibitem{Kleeberg:2016}
K.~Kleeberg, D.~Smith, L.~von Smekal, in preparation  (2016).

\bibitem{Assaad:13:1}
F.~F. Assaad, I.~F. Herbut, Phys. Rev. X {\bf 3} (2013), 031010,
  \href{http://arxiv.org/abs/1304.6340}{ArXiv:1304.6340}.

\bibitem{Buividovich:16:1}
P.~V. Buividovich, M.~V. Ulybyshev, Int. J. Mod. Phys. A {\bf 31} (2016),
  1643008, \href{http://arxiv.org/abs/1602.08431}{ArXiv:1602.08431}.

\bibitem{Herbut:14:1}
M.~Hohenadler, F.~{Parisen Toldin}, I.~F. Herbut, F.~F. Assaad, \href{http://dx.doi.org/10.1103/PhysRevB.90.085146}{Phys. Rev. B
  {\bf 90} (2014), 085146}.%, \href{http://arxiv.org/abs/1407.2708}{ArXiv:1407.2708}.

\end{thebibliography}

\end{document}